\documentstyle[prl,aps,psfig]{revtex}
\begin{document}
\draft

\title{Modulation of the dephasing time for a
magnetoplasma in a quantum well}
\author{M. W. Wu and H. Haug}
\address{Institut f\"ur Theoretische Physik, J.W. Goethe Universit\"at
Frankfurt, Robert-Mayer-Stra\ss e 8, D-60054 Frankfurt a. M.,
Germany}
\date{\today; E-mail: wu@mandala.th.physik.uni-frankfurt.de}
\maketitle
\begin{abstract}
We investigate the femtosecond kinetics of
optically excited 2D magneto-plasma. We calculate the femtosecond dephasing
and relaxation kinetics of the laser
pulse excited magneto-plasma due to bare Coulomb potential scattering,
because screening is under these conditions of minor importance.
By taking into account four Landau subbands in both the conduction
band and the valence band, we are now able to extend our earlier
study [Phys. Rev. B {\bf 58}, 1998,in print] to lower magnetic fields. We can
also 
fix the magnetic field and change the detuning to further investigate
the carrier density-dependence of the dephasing time. For both cases, we
predict strong modulation in the dephasing time.
\end{abstract}

\pacs{Keywords: A. quantum wells, A. semiconductors, D. electron-electron interactions,
D. optical properties}

Numerous experimental and theoretical
studies have been devoted to the problem of transient charge fluctuations
induced by femtosecond pulse excitation in semiconductors
which  can be studied through
nonlinear-optical effects to elucidate many-body phenomena,
such as time-dependent Coulomb correlations.
Most of the experimental studies have been performed without
magnetic field\cite{proce,shah,haug}. The few
femtosecond optical studies in the presence of a
strong magnetic field focused on low-density
magneto-excitons\cite{stafford,glut,siegner,wegener}. With strong
resonant laser
pulses which excite a dense carrier system---with a density
above the Mott ionization density---in a strong magnetic
field, one can study the relaxation and dephasing kinetics
of a magnetoplasma. This problem becomes important as experimental studies
of the relaxation and dephasing kinetics in QW's and superlattices
are in progress\cite{private}.

Recently, we presented a first kinetic study for a femtosecond
laser-pulse excited 2D dense non-equilibrium magnetoplasma in a QW
in the framework of the semiconductor Bloch equations combined
with Coulomb scattering rates\cite{wu}. We assumed an additional weak
lateral confinement which lifts the degeneracy of the Landau
levels partially. We expanded the density matrix of
a two-band ({\em i.e.}, the conduction band and the valence band)
semiconductor in the eigenfunctions of the 2D electron in the
presence of the strong magnetic field and the weak parabolic confinement.
We formulated the scattering terms for the population
distribution functions of
the various Landau subbands and for the optically induced polarization
components between the Landau-subbands in the conduction and valence band in
the form of non-Markovian quantum kinetic scattering integrals\cite{haug} 
and in the form of semiclassical Boltzmann-type scattering rates.
We calculated the time-resolved (TR)
and time-integrated (TI) four-wave mixing (FWM) signals for two 50 fs pulses
by taking into account up to three Landau subbands in both the valance
band and the conduction band. The carrier frequency of the
two delayed pulses is tuned slightly above the unrenormalized energy gap.
We simplified the problem by assuming equal
effective electron and hole masses, as it can be approximately realized in
strained QW's. Naturally, unequal effective masses will
lead to more complicated  quantum beat structures in the FWM signals
and modify to some extent also the resulting
relaxation and dephasing rates. Thus our studies should be seen
only as an idealized model calculations.
Bare Coulomb potential is used in our calculation
because screening is under these strong confinement conditions of minor
importance (see e.g. Ref.\onlinecite{HaugKoch}). Naturally in the limit of
vanishing magnetic field a Boltzmann kinetics with a bare Coulomb potential is
not justified.

We find in our preceding paper\cite{wu} that the FWM signals exhibit quantum
beats mainly with twice the cyclotron frequency.
Contrary to general expectations, we find no pronounced slowing down
of the dephasing with increasing magnetic field.
On the contrary, one obtains in same ranges of the magnetic field a
decreasing dephasing time because of the increase of the Coulomb matrix
elements and the number of states in a given Landau subband. In the
situation when the loss of scattering channels exceeds these increasing
effects, one gets a slight increase of the dephasing time.
However, details of the strongly modulated scattering kinetics depend
sensitively on the detuning, the plasma density, and the spectral pulse width
relative to the cyclotron frequency.  

As discussed in our previous paper, we took only three Landau subbands
in our calculation. This is mainly because of the expansion of
number of matrix elements of Coulomb scattering
increases as $N^4$ with $N$ being the total
number of Landau subbands considered. With $N=3$ in our previous calculation,
the number of form factors is already 81. However, such low number of
Landau subbands limits us to the magnetic fields higher than 10\ T.
More Landau subbands are necessary in order to extend the kinetics
to lower magnetic fields or for larger detunings.

In this report, we take into account four Landau subbands in both
the conduction band and the valence band.
256 matrix elements of Coulomb scattering are calculated in the same way as
discussed in the Appendix of Ref.\ \onlinecite{wu}. So we can investigate
the femtosecond dephasing and relaxation kinetics of magnetoplasma
with magnetic field $B>6$\ T. We can also fix the magnetic field
and tune the laser pulses over a few Landau subband transitions.
We find strong modulation of the dephasing time both for
variations of the magnetic field and the detuning.
These modulations could not be seen fully in our
previous paper due to the small range of available $B$ fields limited by
three considered Landau subbands.

The semiconductor Bloch equations are all the same as those in our previous
paper\cite{wu}:
\begin{equation}
\label{eqs}
\dot \rho_{\nu ,n,\nu^\prime,n^\prime,k}=\left.\dot
\rho_{\nu ,n,\nu^\prime,n^\prime,k}\right |_{\mbox{coh}}
+\left.\dot \rho_{\nu ,n,\nu^\prime,n^\prime,k}\right |_{\mbox{scatt}},
\end{equation}
with $\rho_{\nu ,n,\nu^\prime,n^\prime,k}$ representing the
single-particle density matrix with the band indices $\{\nu,\nu^\prime
\}=\{c,v\}$ and the correspondung Landau subbands $\{n,n^ \prime\}$.
The diagonal elements describe the carrier distribution functions
$\rho_{\nu ,n,\nu ,n,k}= f_{\nu nk}$ of the $n$-th
Landau subband and the wavevector $k$, and the
off-diagonal elements describe the interband polarization components, e.g.
$\rho_{c,n,v,n,k} = P_{nk}e^{-i\omega t}$.  For the
assumed e-h symmetry, $f_{enk}\equiv
f_{hnk}\equiv f_{nk}$ and the polarization has only components between
subbands of the same quantum number n in the conduction and valence band,
which simplifies the problem considerably. The coherent parts of the
equations of motion for the distribution functions and the polarization
components include Hartree-Fock contributions and can be found
in our previous paper. So can the explicit forms of the
scattering rates. However in this report, we
only take the Markovian limit. The Landau index $n$ in our
present study ranges from 0 to 3.

We use the same material parameters of the quantum well as our
previous paper. We perform a numerical study of the Bloch equations in
the Boltzmann limit to calculate TR and TI FWM signals
in order to study the effective dephasing time.
To do so, we use two delayed Gaussian pulses of a width of 50 fs
and a variable delay time $\tau$,
$E_0(t)=E_0(t)+E_0(t-\tau)e^{i\varphi}$ with
the relative phase $\varphi=({\bf k}_2-{\bf k}_1)\cdot {\bf x}$
resulting from the different propagation directions
${\bf k}_1$ and ${\bf k}_2$.
We use an adiabatic projection technique with respect to this phase
in order to calculate
the polarization in the FWM direction with wavevector $2{\bf k}_2
-{\bf k}_1$ described in detail
in Ref.\ \onlinecite{banyai}.  This technique is suitable for optically
thin crystals, where the spatial dependence can be treated
adiabatically\cite{koch}.
The intensity of each
pulse is given by $\int_{-\infty}^\infty dE_0(t)dt=\chi\pi$ with
$\chi$ denoting the fraction of a $\pi$-pulse defined without local
field corrections and $d$ being the optical-dipole matrix element.
Differing from our previous paper where we discussed
both the intermediate density case ($\chi=0.1$) and high density
case ($\chi=0.3$), in this study we focus only on the intermediate
density case with $\chi$ fixed to 0.1.

Our main results are plotted in Figs. 1 and 3. In Fig. 1 we plot the
effective dephasing time as function of magnetic field $B$ for
pulses with detuning $\Delta_0=26.4$\ meV, which is the same
value used in our previous calculation\cite{wu}.  The effective dephasing time
$T_{eff}$ is obtained from the decay of the TI-FWM signal with the delay time
$\tau$ written in the form $\propto \exp(-\tau /T_{eff})$.
 The solid curve is
our present calculation with 4 Landau subbands and the dashed curve
is our earlier one with 3 Landau subbands. We found they coincide
above $B=15$\ T which is in agreement with our discussion that
3 Landau subbands are only good for high magnetic fields and
one needs to include more Landau subbands for lower magnetic fields.
We further find the modulation of dephasing time as it first
decreases with decreasing magnetic field and increases again when
$B$ decreases from 7\ T to 6\ T. We speculate that
more modulations occur as the magnetic field is still lower.
However, we cannot push our calculation to lower fields because that
would require even more Landau subbands.

This modulation can be well understood in the
way of our previous discussion\cite{wu}. For fixed pulses,
several effects compete with each other
when the magnetic field increases. On one hand,
the number of Landau subbands which contribute to the Coulomb scattering
kinetics decreases. In particular the contributions to the dephasing from the
intra- and inter-subband scattering of the higher Landau subbands as well as
the inter-subband scattering between the higher and lower subbands decrease.
For large populations in one subband the Pauli blocking may further
reduce also the intra-band scattering rates. All these effects (we refer
to them as effects I in the following)
increase the dephasing time. On the other hand, with increasing $B$
field the degeneracy of Landau subbands increases and the matrix elements of
the Coulomb scattering become larger. Moreover, an increasing
degeneracy also increases the scattering rates.
Both the increased degeneracy and the increased Coulomb matrix
elements (effects II) reduce the dephasing time. When effects I dominate
over effects II, one observes increase of the dephasing time. Otherwise,
a decrease of the dephasing time results.

In order to further understand the properties of dephasing, we
change the detuning for a fixed magnetic field of $B=7$\ T.
In Fig.\ 2 we illustrate the pulse spectra tuned
at $-20$\ meV which is far below the band gap (solid curve) and
at 40\ meV which is deep inside the band (dashed curve).  We tune
the laser pulses from around $-35$\ meV to 40 meV and calculate the
effective dephasing time. The resulting dephasing time is plotted in Fig.\ 3
as function of the detuning $\Delta_0$. From Fig.\ 3 one can see strong
modulations of the dephasing time. When the laser pulses are tuned
far away from the lowest optical transition, the excitation is very small
and the dephasing time is independent on the detuning (and the related
carrier densities). However the
detuning strongly affects the dephasing time when it is larger than
0. We find the dephasing time reaches minima when the laser
pulses are tuned resonantly at $P_0$ and $P_1$. Another minimum
is observed when $\Delta_0$ sits between $P_0$ and $P_1$ and
the pulse excites comparable populations in both bands.

For fixed magnetic field, the matrix elements of Coulomb scattering
are fixed. The dephasing time is mainly modulated by the occupation
of Landau subbands with varying detuning of the laser pulse.
When the pulses are resonant with an optical transition, the
carriers mainly populate the corresponding Landau subband. This
makes the Coulomb scattering more efficient as the scattering
rate increases superlinearly with the carrier density.
Therefore one observes
a minimum in the dephasing time. It is noted here that the distribution
function in this calculation is smaller than 0.5 even after the second
pulse. This rules out a dominant contribution of
Pauli blocking which makes the
dephasing times longer as discussed before. The third minimum
between $P_0$ and $P_1$ in
Fig.\ 3 comes from the fact that the center of the pulse sits
just between the two lowest optical transitions and it
pumps carriers with both of its tails. In this situation
the lowest two Landau subbands both get relatively
large excitations and therefore lead to the fast dephasing.
It is noted that not withstanding the fact that
we plotted the detuning to $-35$\ meV, the physically meaningful
range is only $\Delta_0>0$, where the carrier-carrier scattering
dominates over other processes.

In conclusion, we have discovered modulations of the effective dephasing time
of 2D magnetoplasma by either fixing the detuning and changing
the magnetic field or fixing the magnetic field and changing the detuning.

We acknowledge financial support by the DFG within the
DFG-Schwerpunkt ``Quantenkoh\"arenz in Halbleiter''.
Interesting discussions with D.S. Chemla and H. Roskos are
appreciated.

\begin{figure}[htb]
\psfig{figure=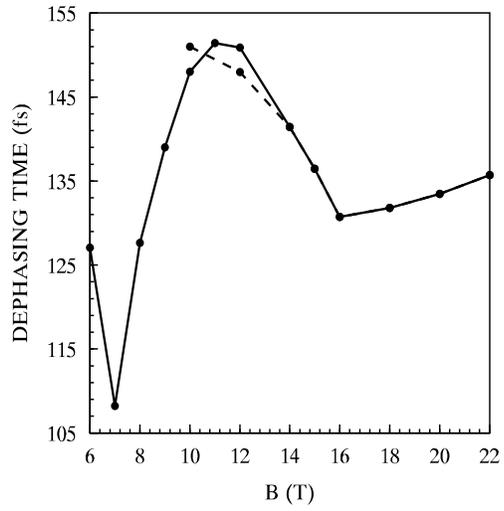,width=8.5cm,height=7.5cm,angle=0}
\caption{Dephasing time as a function of $B$ for fixed
detuning $\Delta_0=26.4$\ meV. The dashed curve  gives the result of
Ref. 10 calculated with only three pairs of Landau subbands.}
\end{figure}
\newpage

\begin{figure}[htb]
\psfig{figure=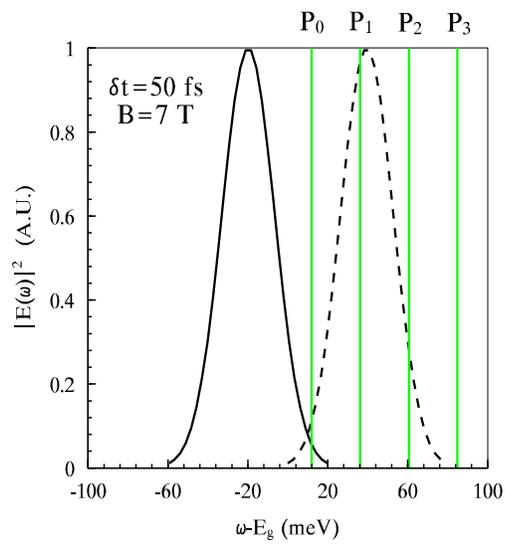,width=8.5cm,height=7.5cm,angle=0}
\caption{The pulse intensity spectrum $|E(\omega)|^2$
for a 50\ fs pulse tuned at $-20$\ meV (solid curve) and
40\ meV (dashed curve), together with the unrenormalized
band edges (with the label $P_n$) for the optical transitions between the
Landau subbands $n$(=0, 1, 2, 3) plotted as solid lines for $B=7$\ T.}
\end{figure}
\newpage

\begin{figure}[htb]
\psfig{figure=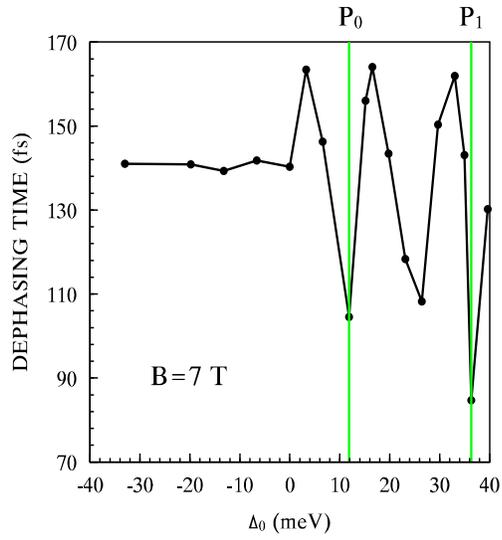,width=8.5cm,height=7.5cm,angle=0}
\caption{Dephasing time as a function of detuning $\Delta_0$ for
$B=7$\ T. The solid lines are the unrenormalized band edges for the
optical transition between the lowest two Landau subbands $n$(=0 and 1).}
\end{figure}

\end{document}